\documentclass[a4paper,showkeys,floatfix,aps,pre,reprint,groupedaddress]{revtex4-1}
\usepackage{mathtools}
\usepackage{amsmath,amssymb}
\usepackage{graphics,graphicx}
\usepackage{dcolumn,bm}
\usepackage{psfrag}
\usepackage{color}
\begin{document}

\title{Prediction of depinning transitions in interface models using Gini and Kolkata indices}

\author{Diksha}
\affiliation{Department of Physics, SRM University - AP, Andhra Pradesh 522240, India}

\author{Gunnemeda Eswar}
\affiliation{Department of Physics, SRM University - AP, Andhra Pradesh 522240, India}

\author{Soumyajyoti Biswas}
\affiliation{Department of Physics, SRM University - AP, Andhra Pradesh 522240, India}

\begin{abstract}
The dynamics of driven interfaces through disordered media is a common framework for a myriad of diverse systems starting from mode-I fracture, vortex lines in superconductors, magnetic domain walls to invading fluid in a porous medium to name a few. The dynamics, for a slow enough drive, progress through punctuated equilibria, showing an intermittent, scale free response statistics that are characterized by only a few parameters. For a strong enough driving, on the other hand, the interface is depinnind i.e., can move continuously through the embedded medium. This depinning transition, often characterized as a critical phenomenon, is a crucial point for the system in question, since it often represent a drastic change in the system (catastrophic failure in the case of fracture, magnetisation switching in the case of domain walls, breakthrough point in the case of fluid invasion etc.). In this work, we outline a framework that can give a precursory signal of the imminent depinning transition by monitoring the variations in sizes or the inequality of the intermittent responses of a system seen prior to the depinning point. In particular, we use measures traditionally used to quantify economic inequality, the Gini index  and the Kolkata index, for the case of the unequal responses of pre-critical systems. The crossing point of these two indices serves as a precursor to imminent depinning. Given a scale free size distributions of the responses, we calculate the expressions for these indices, evaluate their crossing points and give a recipe for forecasting depinning transitions. We compare these with simulation results from the Edwards-Wilkindon, KPZ and fiber bundle model interface with variable interaction strength. The results are applicable for any interface dynamics undergoing depinning transition. The results also explain previously observed near-universal values of Gini and Kolkata indices in self-organized critical systems.
\end{abstract}

\maketitle
\section{Introduction}
An interface driven through a disordered medium is a situation that appears in a wide variety of physical systems, such as domain wall dynamics through a disordered magnet \cite{zapperi98,domain}, fluid front invading a disordered porous medium \cite{Porous,Porous1}, vortex lines in superconductors \cite{larkin79,Huse}, fracture front in mode-I fracture \cite{mode-I,mode1} to name a few. The interface is acted upon by an external driving force,  the different parts of the interface interact with an `elastic' force, which acts like a surface tension, and finally the path of the interface faces pinning centers that try to block the particular parts of the interface from moving forward through the medium. The competing nature of the driving and pinning forces implies that there is a critical value for the driving force for a particular distribution of the pinning forces, beyond which the interface will keep on moving forward, overcoming all pinning centers on its way. For driving force values smaller than the critical value, the interface will be pinned i.e., will be arrested by the pinning centers. 

The critical value of the driving force needed for the depinning transition is non-trivial, given that the dynamics of the interface is a cooperative phenomenon, mediated by the `elastic' interacting between the parts of the interface.
The transition between pinned and moving phases of the interface is, therefore, a second-order, non-equilibrium critical phenomenon. The tools of critical phenomena, namely the universality hypothesis, therefore apply, which can help in estimating the critical exponent values \cite{stanley}. However, given the nature of the systems where depinning can occur, it is often also vital to have an estimate for the distance from the imminent transition point, when such a point is approached from the sub-critical regime. For example, if the mode-I fracture \cite{mode-I} is studied for a disordered sample, then a depinning transition would imply a catastrophic failure of the sample. Similarly, if the dynamics of a magnetic domain wall \cite{zapperi98} is studied using an external magnetic field, a depinning transition would mean a `catastrophic' switching of the magnetic state of the sample (say, from majority up spin to majority down spin) \cite{das}.  In the case of fluid flow through porous media \cite{Porous}, depinning would result in a break-through of the invading fluid through the medium in question. 

Therefore, much effort has been spent in finding ways to forecast imminent depinning transition from the statistical regularities in the response of a system approaching such a transition \cite{alava,karppinen,graham,saichev,kun,cubuk,bapst,Papani,creep}. As mentioned above, depinning transitions are often characterized as (non-equilibrium) critical phenomena, which results in having the possibility of a growing correlation length within the system that in-turn causes cooperative response statistics. Keeping track of this growing correlation, therefore, is a rather non-invasive and optimal monitoring pathway. Specifically, it is known that the dynamics of a system near the depinning transition start showing intermittent behavior. For example, in fracture front propagation, this would be intermittent acoustic energy emissions that have scale-free size distributions \cite{bonamy}. For fluid invasion, the changes in the invaded volume with time would show similar intermittent scale-free statistics (see e.g., \cite{toussaint}). A similar observation exists for changes in magnetization for magnetic domain wall depinning \cite{sethna}. It is natural to expect that these intermittent dynamics, together with the scale-free statistics of their `sizes', would mirror the `health' of the system in terms of their stability or the proximity to a catastrophic (depinning) event. 

In recent years, some attempts were made to locate the critical point in a system by quantifying how much unequal the responses of a system are as the system approaches the critical point \cite{diksha}. There are several different ways to locate (or forecast) the critical point using this method (see e.g., \cite{soc}). In all cases, however, the quantification of inequality (or variation) in the `sizes' of response (for example, the unequal avalanche sizes for fracture) is a central requirement. In order to do this quantification, we borrow the idea of inequality indices that are traditionally used in socio-economic systems, namely the Lorenz curve and the inequality indices derived from it. For our purposes here, we will use two such quantities: the Gini index ($g$) and the Kolkata index ($k$). 

As can be intuitively understood, the avalanche sizes in the case of (quasi-brittle) fracture gradually become more unequal, as the failure point approaches. This is generally true for depinning transitions. In analogy to critical phenomena, this can be attributed to a growing correlation within the system close to the critical point. Therefore, the inequality measures would also grow near the critical point. There are universal scaling properties for the functional forms of $g$ and $k$, which can then be used to estimate the critical points and scaling relations in such systems \cite{diksha1}. Particularly, the values of $g$ and $k$, when measured for the response functions of a system (for example, the growing avalanche sizes in fracture), cross each other slightly ahead of the critical point. The crossing of these two indices, therefore, could be used as a precursor to the approaching critical point in those systems (see e.g., \cite{soumyaditya,manna}). It is important to note here that in the cases where the critical point has already been reached (for example, the self-organized critical systems), the values of $g$ and $k$ continue to remain close to each other. This have been observed extensively in analyzing data for several socio-economic systems, which were conjectured to be in the self-organized critical state \cite{soc}.  

In this work, we use the inequality indices to detect the depinning transition point in elastic interfaces driven through disordered media. Particularly, we study the Edwards-Wilkinson, KPZ and fiber bundle interfaces in the pre-critical state. From the time series of avalanches, we formulate a framework for having precursory signals of imminent failures.

\begin{figure*}
    \includegraphics[width=17cm]{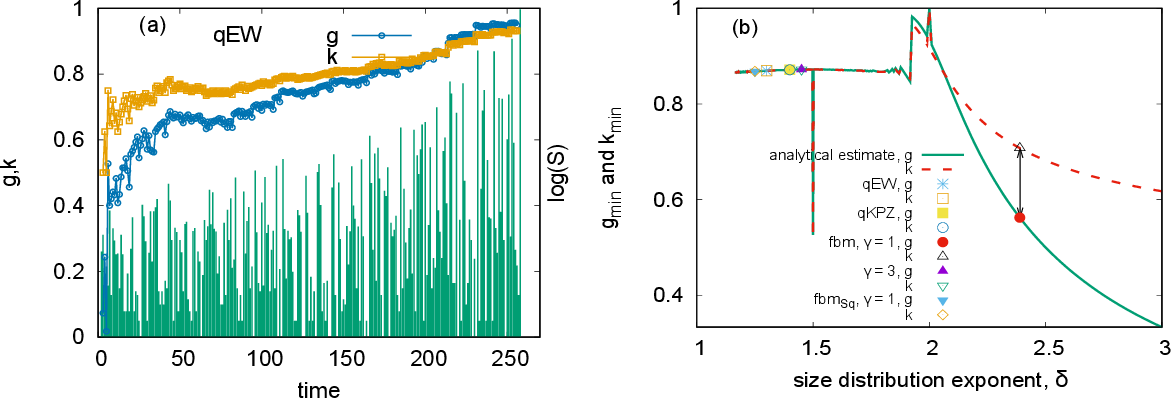}
    \caption{(a) The time series of log of avalanche series is shown along with $g$ and $k$ for the quenched Edwards-Wilkinson Model. Here crossing of $g$ \& $k$ occurs prior to depinning transition. Therefore, the crossing of $g$ \& $k$ can be a very good indicator before the depinning transition. (b) $g_{min}$ and $k_{min}$ are the values where $g$ \& $k$ come closest to each other except at 1.5. As can be seen, for $\delta>2$, they do not cross. For $\delta<2$, $g_{min}$=$k_{min}<1$, which means $g$ \& $k$ cross each other for qEW, qKPZ, and Fiber bundle interface model.}
    \label{gk_min}
\end{figure*}

\section{Quantification of inequality of avalanches: The inequality indices} 
Inequality in the distribution of resources, such as income or wealth, is a common issue in societies. Therefore, the quantification of inequality was a natural question in social sciences. Over the years, several indices were proposed to quantify inequality. For instance,  the Gini index ($g$) \cite{gini} (which assesses wealth inequality), and the more recently introduced Kolkata index ($k$) \cite{kolkata} among others, are used for such quantification. Both of these inequalities are quantified through the Lorenz curve $\mathcal{L}(p)$ \cite{lorenz}. The Lorenz curve denotes that the $p$ smallest fraction of the events accounts for $\mathcal{L}(p)$ fraction of the total events. In terms of wealth, the poorest $p$ fraction of the individuals own $\mathcal{L}(p)$ fraction of the total wealth. Along the same line, the definition could be extended to other (non-negative) quantities, such as an avalanche time series, where the Lorenz curve would denote $p$ fraction of the smallest avalanches accounting for $\mathcal{L}(p)$ fraction of the total avalanche mass. 

The two inequality indices $g$ and $k$ are then defined through the Lorenz function \cite{lorenz}: with $g=1-2\int\limits_0^1\mathcal{L}(p)dp$ and solving the fixed point $1-k=\mathcal{L}(k)$ respectively (see Appendix for a detailed discussion). The Lorenz curve $\mathcal{L}(p)$ is typically represented as a nonlinear curve that always lies below the line of perfect equality (diagonal from (0,0) to (1,1)), which represents a situation where all avalanches (or events) have equal size. The Gini index is the area between the equality line and the Lorenz curve, normalized by the area under the equality line ($1/2$). The range of $g$ varies between $0$ to $1$ where $g=0$ corresponds to perfect equality (all sizes equal) and $g=1$ corresponds to extreme inequality (all but one size non-zero). The Kolkata index says that the $1-k$ fraction of the largest avalanche accounts for the $k$ fraction of total damages (avalanche mass), and the range of $k$ varies between $0.5$ to $1$ where $k=0.5$ implies complete equality and $k=1$ indicates the extreme inequality. This is a generalization of the Pareto's 80-20 law \cite{pareto}.

In recent years, these inequality measures have been used in physical systems for various different purposes \cite{k_pre,inequality_pre,diksha}. As mentioned before, the critical scaling properties \cite{soumyaditya}, prediction of catastrophic events \cite{diksha} etc. could be studied using these measures.  
In this work, we calculate these inequality measures of time series of avalanche sizes for multiple models of interface depinning in the sub-critical regime. Using the inequalities or variations of sizes between the avalanches in the depinning sub-critical phase of the models, we attempt to infer a precursory signal that can work for driven interfaces with a broad class of interaction kernels.

As indicated above, both the indices ($g$ and $k$) that we are interested in, needs to be evaluated from the Lorenz function $\mathcal{L}(p)$. Now, from a time series with $r_0$ number of terms, of a variable that follows a power law size distribution function (assume $P(S)=CS^{-\delta}$; more realistically a lower and upper cut-offs are necessary), the Lorenz function can be estimated in the following way: First we arrange the series in ascending order. It is easy to then see \cite{lucas} that due to the power-law distribution in sizes, the ordered series will follow a form 
\begin{equation}
S_r=(r-r_0)^{-n},
\label{ordered_series}
\end{equation}
with $n=\frac{1}{\delta-1}$. Of course, the maximum size, occurring at $r=r_0$ is not infinite in physical systems, but limited by a upper cut-off, typically an exponential round-off governed by the system size. 

Now, we are interested in dynamical monitoring of the values of $g$ and $k$, which requires that we only look at a fraction (say, $b$) of the whole time series and evaluate $\mathcal{L}$ (and $g$, $k$) as $b$ goes from 0 to 1. At this point, a crucial observation for further progress in the calculation is that since we are looking at the pre-critical dynamics of the depinning transition, the time series of avalanches is not stationary. Furthermore, given that we are progressively loading the system without dissipation, the avalanche sizes, on average, will continue to increase, as the depinning point is approached. This is captured in Fig.\ \ref{gk_min}(a) notwithstanding the fluctuations, the average size of the avalanches keeps on increasing as the system approaches the depinning point (shown here for the EW model).  Therefore, for practical purposes, it is a fair assumption to make that until any point of time, the actual avalanche series and that obtained after arranging such series in the ascending order, are close to each other (see Appendix Fig. for a comparison). We, therefore, go on to make the estimate of the Lorenz function (and consequently that of $g$ and $k$), from the ordered functional form of Eq. (\ref{ordered_series}).
 We will show later that for our purposes of forecasting imminent transition, this is a fairly good approximation. 

Then, the Lorenz function for a time series up to $b$ fraction (where $b=1$ imply the transition point) can be obtained from:
\begin{equation}
    \mathcal{L}(p,b,n)=\frac{\int\limits_0^{pbr_0} (r-r_0)^{-n} dr}{\int\limits_0^{br_0} (r-r_0)^{-n} dr},
\end{equation}
which takes the form
\begin{equation}
    \mathcal{L}(p,b,n)=\frac{1-(1-pb)^{1-n}}{1-(1-b)^{1-n}}.
\end{equation}
It is then straightforward to evaluate the Gini index as
\begin{equation}
    g(b,n)=1-\frac{2}{1-(1-b)^{1-n}}\left[1+\frac{(1-b)^{2-n}-1}{(2-n)b}\right].
    \label{g}
\end{equation}
The Kolkata index needs to be obtained from the self-consistent solution of
\begin{equation}
    1-k(n,b)=\frac{1-(1-k(n,b)b)^{1-n}}{1-(1-b)^{1-n}}.
    \label{k}
\end{equation}

Remembering that $n=1/(\delta-1)$, with $\delta$ being the size distribution exponent for the avalanches, we can now evaluate $g$ and $k$ as functions of $b$, which is a proxy of time. Therefore, we end up with an estimate of time variation of the inequality indices prior to the depinning transition point. It is then necessary to estimate a precursory signal from the time variation of these two indices. As mentioned before, for sufficiently small values of $\delta$, these two indices would cross each other before depinning (i.e., for $b<1$). In Fig.\ \ref{gk_min}(b) , we show the values of the two indices (denoted by $g_{min}$ and $k_{min}$), when they become closest to each other for the lowest value of $b$ i.e., at the earliest time. This means that if the closest approach of $g$ and $k$ happens for $b<b_0$ and again at $b=b_1$ with $b_0<b_1$, we then keep the values ($g_{min}$ and $k_{min}$) for $b_0$. Now, it is seen that for $\delta>2$, $g$ and $k$ do not become equal ($g_{min}\ne k_{min}$) i.e., there is a finite gap even at $b=1$. For other values of $\delta$, except at some special cases, the two indices become equal for $b<1$. The equality of $g$ and $k$ is then a precursory signal of an imminent depinning transition point. 

We have indicated a few points in the plots that correspond to the expected values of $g_{min}$ and $k_{min}$ for different models. For the cases where the avalanche size distribution exponent is steeper than  -2, there will be no crossing of $g$ and $k$. But one can then consider the square of the avalanche sizes, which will have a shallower size distribution and will therefore have a crossing point. We will discuss the relevant cases of the individual models in the subsequent sections. 

One final point to note before moving to the individual models is the near-constant values of $g_{min}$ and $k_{min}$ for different values of $\delta<2$ (excluding $\delta=1.5$). This range of $\delta$ values are observed in a very wide range of physical and socio-economic systems, especially in the context where they are stipulated to be in a self-organized critical (SOC) state. A measure of $g$ and $k$ for such systems, using a reasonable long-time series that will be in a stationary state, will yield values close to $0.87$, as indicated in Fig.\ \ref{gk}. This is precisely what was observed in data quite extensively \cite{soc,kolkata} (see also \cite{zoltan,cscore} for citation of scientists) and also in simulations (see e.g., \cite{manna}). However, our focus here is on pre-critical time series, which follow power law size distributions, but unlike SOC, are non-stationary series. 

\begin{figure}
    \includegraphics[width=8cm]{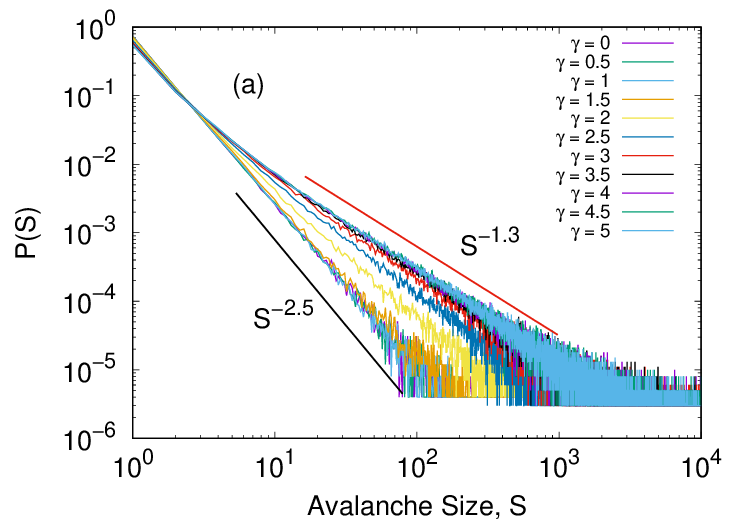}
    \includegraphics[width=8cm]{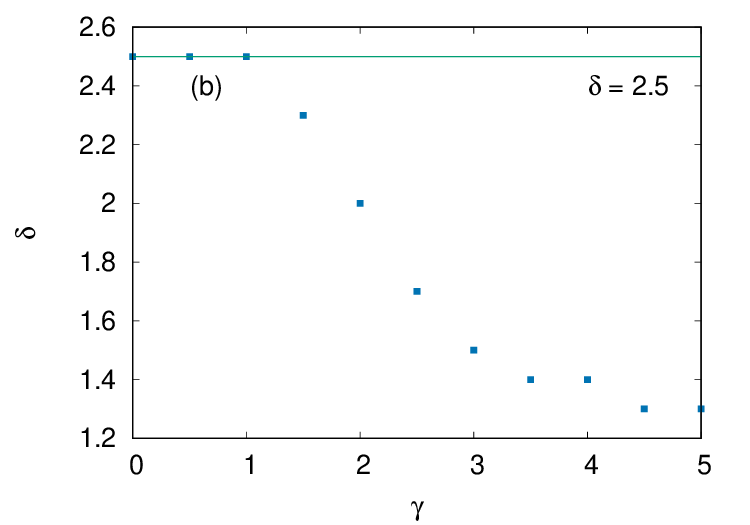}
    \caption{The avalanche size distribution $P(S)$ for the Fiber bundle interface model are plotted for the different values of $\gamma$ = $0$, $0.5$, $1$, $1.5$, $2$, $2.5$, $3$, $3.5$, $4$, $4.5$, \& $5$ for the system size $L$ = 10000 in Figure (a). The variation of the size distribution exponent $\delta$ with $\gamma$ is shown in Figure (b). }
    \label{size_dist}
\end{figure}


\begin{figure*}
    \includegraphics[width=17cm]{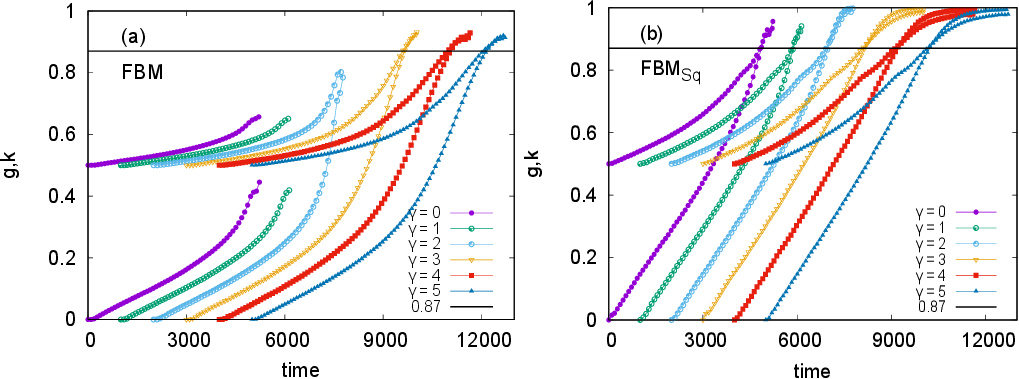}
    \caption{The time variation of $g$ and $k$ are shown for the different values of $\gamma$ = $0$, $1$, $2$, $3$, $4$, and $5$ for the system size $L$ = 10000 in the FBM. Here low $\gamma$ corresponds to the mean field limit and high $\gamma$ corresponds to the localized regime. (a) The crossing of $g$ and $k$ is seen when the $\gamma$ value is high, since the size distribution exponent value is lower there (see Fig. \ref{size_dist}). (b) The crossing of $g$ \& $k$ is always seen for all values of $\gamma$ when we calculate $g$ \& $k$ by taking the square of the avalanche sizes ($S^2$), because the size distribution of $S^2$ will have a smaller exponent value than that of $S$. For visual clarity, all curves, except $\gamma=0$ are shifted in time.}
    \label{gk}
\end{figure*}
 
\section{Models and Methods}
In this work, we consider three models of depinning transition: the fiber bundle interface model \cite{pierce,lucas}, the quenched Edwards-Wilkinson Model \cite{ew}, and the quenched Kardar-Parisi-Zhang model \cite{kpz}. In the fiber bundle interface model, multiple elastic fibers are arranged in a bundle, and each fiber has an associated threshold load that determines its depinning point when subjected to an applied external force.

\subsection{Simulation methods}
Here we outline the details of the simulation methods for different kinds of interfaces that we consider, particularly the fiber bundle interface, and quenched Edwards-Wilkinson and KPZ interfaces. Through intermittent dynamics prior to depininng, the avalanche statistics are recorded and used subsequently to measure the inequality or the variation in their sizes. 

\subsubsection{Simulating FBM}
For the fiber bundle model version of the interface, we consider an array of discrete elements, each having a finite failure threshold ($\sigma_{th}$), drawn from some distribution function. When a load is applied on the interface, some of these elements (fibers) break, and the load carried by those fibers are redistributed among the intact fibers. Those fibers can then break, given that the load level has now increased, triggering an avalanche. The avalanche will subsequently stop when all intact fibers are carrying load that are below their failure thresholds. To model a propagating interface, all fibers broken during the avalanche are then restored with zero initial load and a threshold value for failure drawn from the same distribution function as before. The load on the system is then increased, on all fibers equally, until the next breaking happens and an avalanche starts. The time scale of avalanche is considered to be much smaller than the external loading rate, which enables us to consider a constant value for load during an ongoing avalanche. 

The load redistribution process is considered to be distance dependent i.e., if the $i$-th fiber is broken, the fraction of the load received by the $j$-th fiber is proportional to $\frac{1}{|i-j|^{\gamma}}$, where $\gamma=0$ is the limit of equal load sharing (mean-field) and $\gamma\to\infty$ is the local load sharing limit. In practice, as noted elsewhere \cite{lucas}, $\gamma\approx 3$ is sufficient for local load sharing behavior. 

We simulate for different values of $\gamma$ and record the corresponding avalanches for the system size $L=10000$ and measure the inequalities of such avalanches, as mentioned above.

\subsubsection{Simulating QKPZ and QEW}
In the Edward-Wilkinson equation and Kardar-Parisi-Zhang equation, we simulate the behavior of a one-dimensional interface model with quenched noise. We use these two models to simulate the surface growth and evolution of a discrete interface represented by the height variable $hi(t)$ and the growth rule for the discrete interface model is:
\begin{align}
    h_{i}(t+1) = \begin{dcases*}
        {h_i(t+1)}, & if $ g_i > 0 $,\\
        {h_i(t)}, & otherwise. 
        \end{dcases*}
  \end{align}
  where, 
  \begin{equation}
      g_i=h_{i+1}(t)+h_{i-1}(t)-2h_i(t)+\eta(i,h)+F
  \label{gi_ew}
  \end{equation}
  \begin{equation}
    \begin{multlined}
        g_i=h_{i+1}(t)+h_{i-1}(t)-2h_i(t)+\\
      \frac{1}{2}\left[\frac{h_{i+1}(t)-h_{i-1}(t)}{2}\right]^{2}+
      \eta(i,h)+F
    \end{multlined}
    \label{gi_kpz}
  \end{equation}
  
  where $\eta(i,h)$ is the random delta correlated pinning force and $F$ is the external driving force. Initially height profile of interface $h_i$ is 0 and the periodic boundary conditions are used on the system size $L$. In Eq. (\ref{gi_ew}) and (\ref{gi_kpz}) $g_i$ is the local growth rate at each site of the interface for the EW equation and the KPZ equation respectively. The height of the interface at each site $h_i$ is updated using the Eq. (\ref{gi_ew}) and (\ref{gi_kpz}). The pinning force $\eta$ is selected randomly from a uniform distribution between (-2, 2) in the case of the EW equation and (-5, 5) in the case of the KPZ equation. Any change in that will be mentioned explicitly in the relevant places.

  The simulation proceeds by applying the minimum external driving force on the interface, which is required to move the height of the interface. When the external driving force is smaller than the pinning force, the interface is pinned. When $g_i$ is positive then the height of the interface $h_i$ is moved by 1 and again new pinning forces, drawn from the same distribution, is assigned to the sites. This process continues until no more growth occurs or the local growth rate becomes zero. During this process, the external driving force has remained unchanged, and then it is increased again to restart the dynamics. The number of updated sites in between two stable states is the avalanche size $S$. When the external driving force is stronger than the critical value, the interface moves indefinitely i.e., it is depinned.
  

  \begin{figure*}
    \includegraphics[width=17cm]{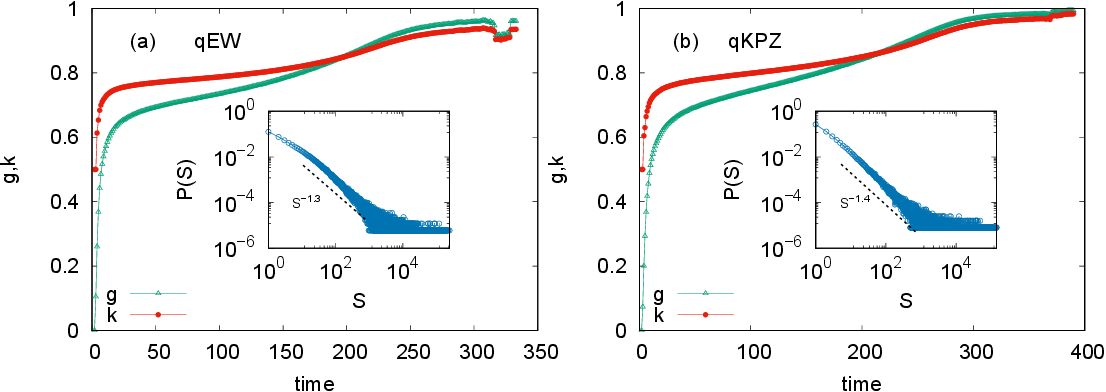}
    \caption{The variation of $g$ and $k$ plotted with time is shown for the two models (Edward-Wilkinson model and Kardar-Parisi-Zhang Model) for the system size $L$ = 1000. In the case of these two models, the crossing of $g$ and $k$ is seen even though if we change the threshold distribution, the depinning is predictable. }
    \label{gk_edw}
\end{figure*}
  
\section{Results}
Here we explain the behavior of inequality indices $g$ \& $k$ of the avalanche time series in the case of EW, KPZ and fiber bundle models. 

In Fig.\ \ref{size_dist}(a) we have plotted the size distribution of avalanche size, denoted as $P(S)$ for the Fiber Bundle Interface model. The size distributions are shown for various values of $\gamma$ including $0$, $0.5$, $1$, $1.5$, $2$, $2.5$, $3.5$, $4$, $4.5$, and $5$. Here low value of $\gamma$ ($<2$) signifies the mean field interaction while a high value of $\gamma$ signifies the nearest neighbor interaction. We observe a continuous variation in the exponent value of the size distribution $\delta$ ranging from $2.5$ to $1.3$ as $\gamma$ varies from $0$ to $5$. In Fig.\ \ref{size_dist}(b) we see that $\gamma = 2$ is the point beyond which the size distribution exponent $\delta$ is below 2. This is in line with earlier results in Ref. \cite{lucas}. The end values of the indices, denoted by $g_f$ and $k_f$, depend on the distribution of avalanche size exponent $\delta$. When these avalanches are arranged in ascending order based on their sizes, the resulting curve will exhibit a divergence characterized by an exponent $n$, where $n=\frac{1}{\delta-1}$  for $n > 1$, $g_f$ and $k_f$ will converge to 1 \cite{soumyaditya}. If the size distribution exponent $\delta$ remains below 2 then the divergence exponent will always be above 1.

Fig.\ \ref{gk}(a) shows the temporal evolution of $g$ \& $k$ for a various $\gamma$ values of including $0$, $1$, $2$, $3$, $4$, and $5$ in the FBM. We observe that when $\gamma$ is low (reflecting a mean-field regime) with a corresponding size distribution exponent $\delta$ of 2.5, $g$ and $k$ do not cross. This implies that in the mean field limit, predicting depinning becomes challenging. However, when $\gamma$ exceeds 2 (indicating the localized regime), the size distribution exponent $\delta$ (magnitude) decreases to 1.3, and the crossing of $g$ and $k$ is seen. In Figure \ref{gk}(b), we square the avalanche size before computing the values of $g$ and $k$. Remarkably, we observe that the crossing of $g$ and $k$ occurs consistently across all values of $\gamma$. As indicated before, the reason is that the size distribution exponent becomes shallower (less than 2) when the avalanche size is squared.

\begin{figure}
    \includegraphics[width=9cm]{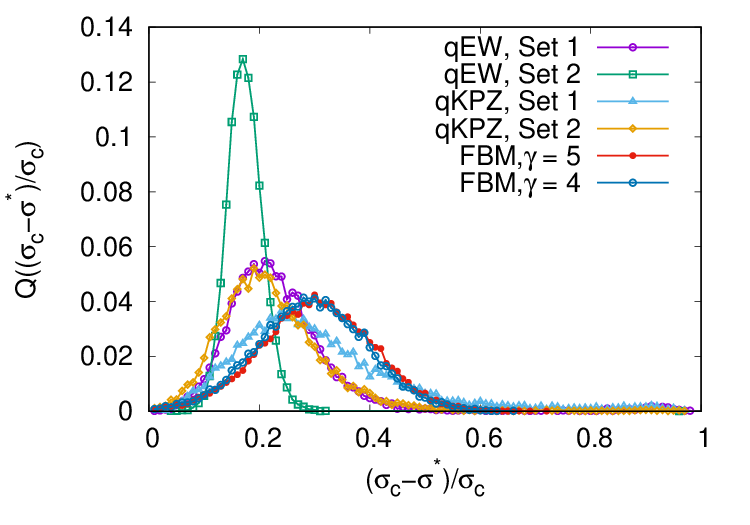}
    \caption{The probability distribution of the $\frac{(\sigma_c-\sigma^*)}{\sigma_c}$ are shown for the three different models (quenched Edwards- Wilkinson model, quenched Kardar-Parisi-Zhang model, and Fiber Bundle Interface Model) for different threshold distributions. Here $\sigma_c$ is the critical load and $\sigma^*$ is the point where $g$ and $k$ cross. In the EW model, we consider two sets, in set $1$ we take the threshold distribution between (-2,2), and in set $2$  we choose the threshold distribution between (-2.5,2.5). In the KPZ model, in set 1 we take the threshold distribution between (-5,5) and in set 2 we take the threshold distribution between (-6,6). In the FBM model, we take the two values of $\gamma$= $4,5$. As can be seen, $g$ and $k$ crosses $10\%$ or $20\%$ away from the depinning and gives the indication of depinning.}
    \label{sigma_dist}
\end{figure}

We extend this analysis to two additional models of depinning transition, namely, the Edwards-Wilkinson (EW) and the Kardar-Parisi-Zhang (KPZ) models. To begin our analysis, we first calculate the size distribution exponent, denoted as $\delta$, for these two models. The size distribution exponent is consistently 1.3 for EW and 1.4 for KPZ (see Fig.\ \ref{gk_edw} inset). Then we calculate the $g$ and $k$ using the time series data of avalanches, and in Fig.\ \ref{gk_edw} we can see the crossing of $g$ and $k$ occurs in both of these cases. This suggests that the crossing of $g$ and $k$ holds significant promise as an effective early indicator preceding the depinning transition in these two models.

Then we calculate the probability distribution of the $Q(\frac{\sigma_c-\sigma^*}{\sigma_c})$ for all three models (FBM, qEW, qKPZ) to check how far the crossing of $g$ and $k$ is happening from the dipinning transition point (see Fig.\ \ref{sigma_dist}). Here $\sigma_c$ represents the critical load at which depinning initiates and $\sigma^*$ signifies the point at which $g$ \& $k$ cross. Here in the fiber bundle interface model, we take the value of $\gamma$ = $4$ \& $5$. In the EW model, we take two sets, in set $1$, we choose a threshold distribution ranging from $(-2,2)$, while in set $2$, we extend the range to $(-2.5,2.5)$. In the KPZ model, for set $1$, we consider the threshold distribution between $(-5,5)$ and for set $2$, we consider it to be $(-6,6)$. In all the cases, we consistently observe that the crossing of $g$ \& $k$ occurs at a distance of about $10\%$ to $20\%$ prior to the depinning transition point, underscoring its robustness as a predictive indicator. All the points of expected $g_{min}$ \& $k_{min}$ values are indicated in Fig.\ \ref{gk_min}(b). It is also to be noted that in no case the x-axis range goes below zero, implying that in every single realisation the crossing happened prior to the depinning.

\section{Discussion and Conclusion}
Forecasting the depinning transition point in interfaces driven through disordered media is an important problem. From catastrophic fracture propagation to magnetization switching, such questions are relevant to physicists and engineers alike. Generally, such cooperative intermittent dynamics leading to catastrophic changes have been subjected to investigations through various rule-based (see e.g., \cite{alava,karppinen,graham,saichev,kun,Papani,diksha1}) and machine-learning  (see e.g., \cite{cubuk,bapst,creep,diksha}) studies. In rule-based investigations, there are often strong sample to sample fluctuations and in cases of machine learning studies, a lack of training set of experiments. 

\begin{figure}
\includegraphics[width=9cm]{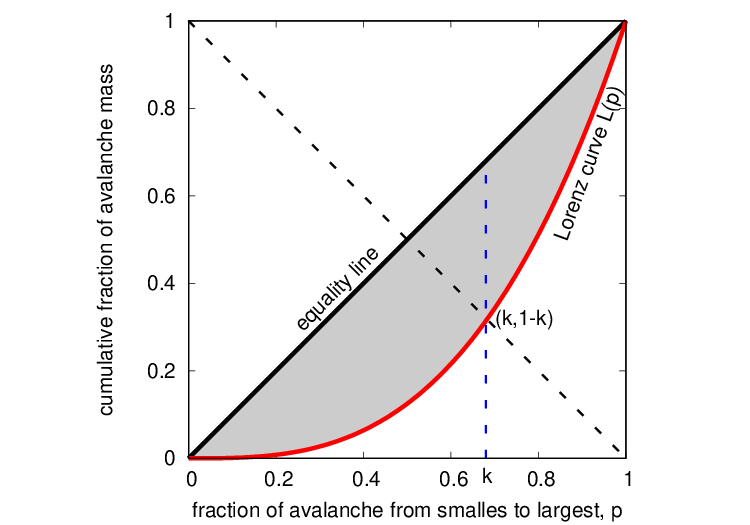}
\caption{A schematic diagram of the Lorenz function $\mathcal{L}(p)$ is shown, where $\mathcal{L}(p)$ denotes the cumulative fraction of the
avalanche mass contained in the smallest $p$ fraction of avalanches. If all avalanches were equal, this would be a diagonal
straight line, called the equality line. The area between the equality line and the Lorenz curve (shaded area), therefore, is a measure
of the inequality in the avalanche sizes. Two quantitative measures of such inequality are extracted from here, the ratio
of the shaded area and that under the equality line (Gini index, $g$) and the crossing point of the opposite diagonal --
from (0,1) to (1,0), shown in the dashed line, and the Lorenz curve, giving the Kolkata index, $k$. $1-k$ fraction of avalanches
contain $k$ fraction of the cumulative avalanche mass.}
\label{fig_s1}
\end{figure}
\begin{figure*}
\includegraphics[width=17cm]{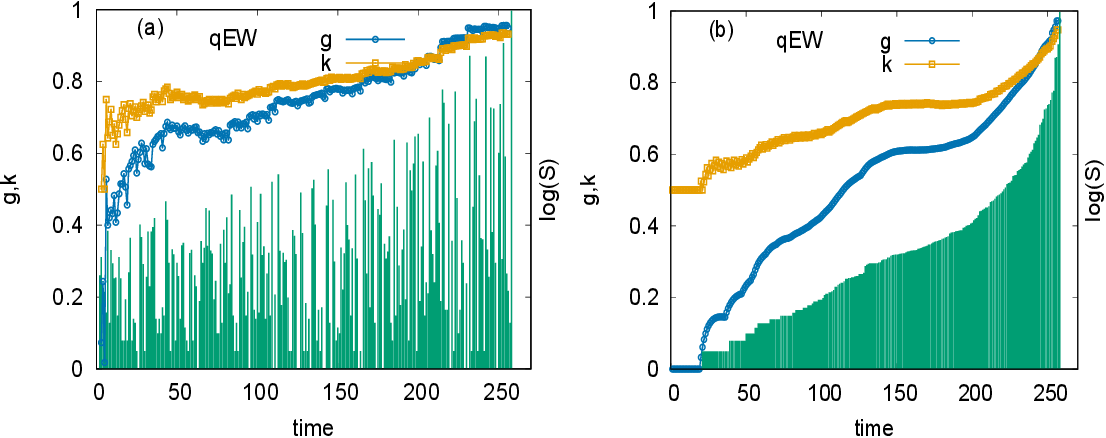}
\caption{(a) time variation of avalanche size along with $g$ and $k$ is plotted for the Edward Wilkinson model. (b) time variation of avalanche size along with $g$ and $k$ is shown when an avalanches are arranged in the ascending order of their sizes.}
\label{fig_s2}
\end{figure*}
In this work, we study this question through the behavior of inequality indices traditionally used in social sciences. The Gini ($g$) and Kolkata ($k$) indices, show remarkably stable characteristics prior to the depinning transition point, when measured using the unequal avalanches of the systems. Recall that the avalanches noted here are not in the stationary state but in the non-stationary pre-critical state, where the average size of the avalanche grows with time (i.e., the applied load, assuming fixed loading rate). Given a power law size distribution of the avalanches, it is possible to calculate approximate expressions for the above two indices (see Eq. (\ref{g}) and (\ref{k}). Then, when the time variation of the indices are looked at, they tend to cross each other at a point prior to the depinning threshold (see Fig.\ \ref{gk_min}(a)). Note that, this phenomenon is not just an average or most probable event, but something that happens on every single realisation of all the models (fiber bundle, qEW, and qKPZ) studied here. Fig.\ \ref{sigma_dist} indeed shows the distribution of the relative distance from the depinning point at which such crossing happened. It is consistently before the critical point in each case. 

As the analytical estimates indicated, for steep enough size distributions (e.g., fiber bundle interface with $\gamma>2$) such crossings do not happen. In that case, however, one can consider the square of the avalanche sizes, which will have a much shallower distribution and will therefore show a crossing in $g$ and $k$. The functional form is hard to estimate since the second moment is not convergent in such cases for fiber bundles. 

Finally, it is worth noting that the value at which the two indices cross, is near-universal (about 0.87) (see Fig.\ \ref{gk}). This is seen from the analytical estimates and also from the numerical simulations of different models. Therefore, the signature of the precursor is not only widely applicable but also near-universal. Indeed, in other contexts such near-universality was noted in simulations and data analysis, which are the results of the size distribution, as noted in Eqs. (\ref{g}) and (\ref{k}). 

In conclusion, we have outlined a framework for forecasting imminent depinning transition in a wide variety of interfaces driven through disordered media, using social inequality indices. The indices, $g$ and $k$, will cross each other at a time prior to the transition point. The analytical estimate and numerical simulations indicate that the precursory signals are near-universal for all such depinning models and is obeyed in every realisation of the simulations performed on Edwards-Wilkinson, KPZ and fiber bundle interfaces. For sufficiently global range of interactions, when $g$ and $k$ of avalanches do not cross, if the indices are measured using the square of the avalanche sizes, then the crossing happens again. This indicates that for all realistic cases, by taking a sufficiently higher power of avalanche sizes, one can use the framework of precursory signals proposed here. An analysis using experimental data could be a fruitful future direction of this study.

\section*{Appendix: Inequality measures for the avalanche time series: Gini ($g$) and Kolkata ($k$) indices}

In this appendix, we explore the characteristics of the avalanche time series of the Fiber Bundle Model (FBM), the quenched Edwards-Wilkinson (qEW) model, and the quenched Kardar-Parisi-Zhang (qKPZ) model.
These time series exhibit a power-law distribution $P(S)\sim S^{-\delta}$, where the probability of large avalanches is few and small avalanches are more common. The exponent value $\delta$ remains consistent across various system sizes and threshold distributions, representing a universal property. However, the critical load required for the system failure is not universal and strongly depends on the details of the system. Recently it has been shown \cite{k_pre} that some time dependent quantities which tend to converge to seemingly universal values near the breakdown point, can be derived from measuring the inequality among avalanche sizes. Historically, some of those are over a century old but used mainly in estimating economic inequality (e.g., Gini index \cite{gini}), and some of them are recently introduced such as the Kolkata index \cite{kolkata}). Monitoring these quantities can serve as valuable indicators for predicting impending failure points. 

In the following section of the appendix, we provide a detailed explanation of how these parameters are defined, how they evolve over time, and how they tend to approach near-universal values as the system approaches the depinning point in the fiber bundle interface, qEW, and qKPZ models.

\subsection*{The Lorenz function, Gini ($g$) and Kolkata ($k$) indices}

In the simulation of the FBM, EW, and KPZ interfaces, the time series of the avalanches can be arranged in the ascending order of their sizes (see Fig.\ \ref{fig_s2}). Then the Lorenz function $\mathcal{L}(p)$ gives the cumulative fraction of the avalanche mass (sum of all avalanche sizes) held by $p$ fraction of the smallest avalanches up to time $t$. It's important to note that if all avalanches were of equal sizes, then the Lorenz function would be a straight line starting from the origin and increasing linearly up to 1. This linear curve is known as the equality line (see Fig. \ref{fig_s1}). However since the avalanches typically have varying sizes, the Lorenz function is non-linear, staying below the equality line and monotonically increasing, with the constraints that $\mathcal{L}(0,t)=0$  and $\mathcal{L}(1,t)=1$. The area between the equality line and the Lorenz function shown as a shaded region in Fig.\ \ref{fig_s1} serves as a measure of the inequality in avalanche sizes. The ratio of this area and that under the equality line ($1/2$ by construction) is called the Gini index $g$ which can be calculated by Eq. \ref{g}.

On the other hand, the ordinate value of the crossing point of the opposite diagonal (the straight line connecting points (0,1) to (1,0)), provides the value of the Kolkata index $k$, which estimates the fraction $1-k$ of the avalanches that collectively account for the fraction $k$ of the total avalanche mass up to that time. The value of the $k$ index is evaluated through Eq. \ref{k}. 

As can be seen from Fig. \ref{fig_s2}, when the avalanches are kept in their order of occurrence and when they are arranged in the ascending order of their sizes, they do not vary too much. Particularly, the crossing point of $g$ and $k$ are similar. Therefore, the approximation made in the calculations when the avalanches are assumed to be in the ascending order, is good enough.

\section*{acknowledgement} 
 The simulations were done using HPCC Surya in SRM University - AP. The authors are thankful to Bikas K. Chakrabarti for useful comments on the manuscript.


\begin{thebibliography}{99}



\bibitem{zapperi98}
S. Zapperi, P. Cizeau, G. Durin, H. E. Stanley,
{\it Dynamics of a ferromagnetic domain wall: Avalanches, depinning transition, and the Barkhausen effect},
Phys. Rev. B {\bf 58}, 6353 (1998).

\bibitem{domain}
David A. Huse and Christopher L. Henley,
{\it Pinning and Roughening of Domain Walls in Ising Systems Due to Random Impurities},
Phys. Rev. Lett. {\bf 54}, 2708 (1985).

\bibitem{Porous}
M. Cieplak and M. O. Robbins,
{\it Dynamical Transition in Quasistatic Fluid Invasion in Porous Media},
Phys. Rev. Lett. {\bf 60}, 2042 (1988).

\bibitem{Porous1}
M. A. Rubio, C. A. Edwards, A. Dougherty, and J. P. Gollub,
{\it Self-affine fractal interfaces from immiscible displacement in porous media},
Phys. Rev. Lett. {\bf 63}, 1685 (1989).

\bibitem{larkin79}
A. I. Larkin, Yu. N. Ovchinnikov,
{\it Pinning in type II superconductors},
J. Low Temp. Phys. {\bf 34}, 409 (1979).

\bibitem{Huse}
Daniel S. Fisher, Matthew P. A. Fisher, and David A. Huse,
{\it Thermal fluctuations, quenched disorder, phase transitions, and transport in type-II superconductors},
Phys. Rev. B {\bf 43}, 130 (1991)

\bibitem{mode-I}
J. P. Bouchaud, E. Bouchaud, G. Lapasset, and J. Planès,{\it Models of fractal cracks},
Phys. Rev. Lett. {\bf 71}, 2240 (1993).

\bibitem{mode1}
L. O. Eastgate, J. P. Sethna, M. Rauscher, T. Cretegny, C.-S. Chen, and C. R. Myers,{\it Fracture in mode I using a conserved phase-field model},
Phys. Rev. E {\bf 65}, 036117 (2002).

\bibitem{stanley}
A. L. Barabási, H. E. Stanley, {\it Fractal Concepts in Surface Growth}, Cambridge University Press 0521483182, p. 366 (1995).

\bibitem{das}
S. Das,  A. Zaig, H. Nhalil, L. Avraham, M. Schultz, L. Klein,
{\it Switching of multi-state magnetic structures via domain wall propagation triggered by spin-orbit torques}, Sci. Rep. {\bf 9}, 20368 (2019).

\bibitem{alava}
J Koivisto, M. Ovaska, A. Miksic, L. Laurson and M. J. Alava,{\it Predicting sample lifetimes in creep fracture of heterogeneous materials}, Phys. Rev. E {\bf 94}, 023002 (2016).

\bibitem{karppinen}
L. Viitanen, M. Ovaska, S. K. Ram, M. J. Alava, and P. Karppinen, {\it Predicting creep failure from cracks in a heterogeneous material using acoustic emission and speckle imaging}, Phys. Rev. Appl. {\bf 11}, 024014 (2019).

\bibitem{graham}
S. Lennartz-Sassinek, I. G. Main, M. Zaiser and C. C. Graham, {\it Acceleration and localization of subcritical crack growth in a natural composite material}, Phys. Rev. E {\bf 90}, 052401 (2014).

\bibitem{saichev}
A. Saichev and D. Sornette, {\it Andrade, Omori, and time-to-failure laws from thermal noise in material rupture}, Phys. Rev. E {\bf 71}, 016608 (2005).

\bibitem{kun}
F. Kun, I. Varga, S. Lennartz-Sassinek and I. G. Main, {\it Approach to failure in porous granular materials under compression}, Phys. Rev. E {\bf 88}, 062207 (2013).

\bibitem{cubuk}
E.D. Cubuk, S.S. Schoenholz, J.M. Rieser, B.D. Malone, J. Rottler, D.J. Durian, E. Kaxiras, and A.J. Liu, {\it Identifying structural flow defects in disordered solids using machine-learning methods}, Phys. Rev. Lett. {\bf 114}, 108001 (2015).

\bibitem{bapst}
Bapst V. et al., {\it Unveiling the predictive power of static structure in glassy systems}, Nat. Phys. {\bf 16}, 448–454 (2020).

\bibitem{Papani}
S Papanikolaou, {\it Microstructural inelastic fingerprints and data-rich predictions of plasticity and damage in solids}, Comp. Mech. 1–14 (2020).

\bibitem{creep}
S Biswas, D Fernandez Castellanos, M Zaiser,
{\it Prediction of creep failure time using machine learning},
Scientific Reports {\bf 10} (1), 16910 (2020).

\bibitem{bonamy}
D. Bonamy and E. Bouchaud, {\it Failure of heterogeneous materials:
A dynamic phase transition?}, Phys. Rep. {\bf 498}, 1 (2011).

\bibitem{toussaint}
Shahar Ben-Zeev, Einat Aharonov, Renaud Toussaint, Stanislav Parez, and Liran Goren, {\it Compaction front and pore fluid pressurization in horizontally shaken drained granular layers},
Phys. Rev. Fluids {\bf 5}, 054301 (2020).

 \bibitem{sethna}
 J. P. Sethna, K. A. Dahmen, and C. R. Myers, {\it Crackling noise},
Nature (London) {\bf 410}, 242 (2001).


\bibitem{diksha}
Diksha, S Biswas,
{\it Prediction of imminent failure using supervised learning in a fiber bundle model},
Physical Review E {\bf 106} (2), 025003 (2022).

\bibitem{soc}
S Banerjee, S Biswas, BK Chakrabarti, SK Challagundla, A Ghosh, SR Guntaka, H Koganti, AR Kondapalli, R Maiti, M Mitra, and Dachepalli R. S. Ram, {\it Evolutionary dynamics of social inequality and coincidence of Gini and Kolkata indices under unrestricted competition},
International Journal of Modern Physics C {\bf 34},2350048 (2023).

\bibitem{diksha1}
Diksha , S Kundu , Bikas K. Chakrabarti, and S Biswas,
{\it Inequality of avalanche sizes in models of fracture},
Physical Review E {\bf 108}, 014103 (2023).

\bibitem{soumyaditya}
S. Das, S. Biswas,
{\it Critical scaling through Gini index}, Phys. Rev. Lett. (in press) arXiv:2211.01281, (2023).

\bibitem{manna}
SS Manna, S Biswas, BK Chakrabarti,
{\it Near universal values of social inequality indices in self-organized critical models},
Physica A: Statistical Mechanics and its Applications {\bf 596}, 127121 (2022).

\bibitem{gini}
C. Gini, {\it Measurement of inequality of incomes}, Economics Journal {\bf 31}, 124126 (1921).

\bibitem{kolkata}
A. Ghosh, N. Chattopadhyay, B. K. Chakrabarti, {\it Inequality in societies, academic
institutions and science journals: Gini and k-indices}, Physica A {\bf 410}, 3034 (2014).

\bibitem{lorenz}
M. O. Lorenz, {\it Methods of Measuring the Concentration
of Wealth}, Publ. Am. Stat. Assoc. {\bf 9}, 209–219 (1905).

\bibitem{pareto}
V. Pareto, A. N. Page, {\it Translation of `Manuale di economia politica' (Manual of political economy)},
A.M. Kelley Publishing, New York (1971).

\bibitem{k_pre}
S. Biswas, B. K. Chakrabarti, {\it Social inequality analysis of fiber bundle model statistics 
and prediction of materials failure}, Phys. Rev. E {\bf 104}, 044308 (2021).

\bibitem{inequality_pre}
Diksha, Sumanta Kundu, Bikas K. Chakrabarti, and Soumyajyoti Biswas,
{\it Inequality of avalanche sizes in models of fracture},
Phys. Rev. E {\bf 108}, 014103 (2023).

\bibitem{zoltan}
T. S. Biró, A. Telcs, M. Józsa, Z. Néda,
{\it Gintropic scaling of scientometric indexes},
Physica A {\bf 618}, 128717 (2023).

\bibitem{cscore}
A. Ghosh, B. K. Chakrabarti, {\it Do Successful Researchers Reach the Self-Organized Critical Point?}, arxiv:2308.14435 (2023).

\bibitem{lucas}
S. Biswas and L. Goehring,
{\it Interface propagation in fiber bundles: local, mean-field and intermediate range-dependent statistics},
New J. Phys. {\bf 18}, 103048 (2016).

\bibitem{pierce}
F. T. Pierce, {\it 32—X.—Tensile tests for cotton yarns v.—“The
weakest link” theorems on the strength of long and of composite
specimens}, J. Textile Inst. Trans. {\bf 17}, T355 (1926).


\bibitem{ew}
J. M. Kim, H. Choi, {\it Depinning Transition of the Quenched Edwards-Wilkinson Equation}, Journal of the Korean Physical Society, {\bf 48}, S241 (2006).

\bibitem{kpz}
H. S. Song and J. M. Kim,
{\it A Discrete Model of the Quenched Kardar-Parisi-Zhang Equation},
Journal of the Korean Physical Society, {\bf 51}, (2007).



\end{thebibliography}

\end{document}